\documentclass[letter,longauth,traditabstract]{aa} 

\usepackage{amsmath}
\usepackage{txfonts}
\usepackage{graphicx}
\usepackage[colorlinks=true, linkcolor=blue, citecolor=blue, urlcolor=blue]{hyperref}
\usepackage{natbib}
\bibpunct{(}{)}{;}{a}{}{,} 
\pdfoutput=1
\begin{document}

\title{The Herschel-ATLAS: The dust energy balance in the edge-on
  spiral galaxy UGC\,4754\thanks{Herschel is an ESA space observatory
    with science instruments provided by European-led principal-
    investigator consortia and with important participation from
    NASA.}}

\author{
M.~Baes\inst{\ref{UGent}}
\and
J.~Fritz\inst{\ref{UGent}}
\and
D.~A.~Gadotti\inst{\ref{ESO}}
\and 
D.~J.~B.~Smith\inst{\ref{Nottingham}}
\and
L.~Dunne\inst{\ref{Nottingham}}
\and 
E.~da Cunha\inst{\ref{Crete}}
\and 
A.~Amblard\inst{\ref{Irvine}}
\and
R.~Auld\inst{\ref{Cardiff}}
\and 
G.~J.~Bendo\inst{\ref{Imperial}}
\and
D.~Bonfield\inst{\ref{Herts}}
\and
D.~Burgarella\inst{\ref{Marseille}}
\and
S.~Buttiglione\inst{\ref{Padova}}
\and
A.~Cava\inst{\ref{IAC},\ref{LaLaguna}}
\and
D.~Clements\inst{\ref{Imperial}}
\and
A.~Cooray\inst{\ref{Irvine}}
\and
A.~Dariush\inst{\ref{Cardiff}}
\and
G.~de Zotti\inst{\ref{Padova},\ref{SISSA}}
\and
S.~Dye\inst{\ref{Cardiff}}
\and
S.~Eales\inst{\ref{Cardiff}}
\and
D.~Frayer\inst{\ref{IPAC}}
\and
J.~Gonzalez-Nuevo\inst{\ref{SISSA}}
\and
D.~Herranz\inst{\ref{Santander}}
\and
E.~Ibar\inst{\ref{ROE}}
\and
R.~Ivison\inst{\ref{ROE}}
\and
G.~Lagache\inst{\ref{IAS-Paris},\ref{Paris-Sud}}
\and
L.~Leeuw\inst{\ref{NASA}}
\and
M.~Lopez-Caniego\inst{\ref{Santander}}
\and
M.~Jarvis\inst{\ref{Herts}}
\and
S.~Maddox\inst{\ref{Nottingham}}
\and
M.~Negrello\inst{\ref{OpenU}}
\and
M.~Micha{\l}owski\inst{\ref{Edinburgh2}}
\and
E.~Pascale\inst{\ref{Cardiff}}
\and
M.~Pohlen\inst{\ref{Cardiff}}
\and
E.~Rigby\inst{\ref{Nottingham}}
\and
G.~Rodighiero\inst{\ref{Padova2}}
\and
S.~Samui\inst{\ref{SISSA}}
\and
S.~Serjeant\inst{\ref{OpenU}}
\and
P.~Temi\inst{\ref{NASA}}
\and
M.~Thompson\inst{\ref{Herts}}
\and
P. van der Werf\inst{\ref{Leiden}}
\and
A.~Verma\inst{\ref{Oxford}}
\and
C.~Vlahakis\inst{\ref{Leiden}}
}

\institute{
Sterrenkundig Observatorium, Universiteit Gent, Krijgslaan 281 S9, 
B-9000 Gent, Belgium
\label{UGent}
\and 
European Southern Observatory, Alonso de Cordova 3107, Vitacura,
Santiago, Chile
\label{ESO} 
\and 
School of Physics and Astronomy, University of Nottingham,
University Park, Nottingham NG7 2RD, UK
\label{Nottingham}
\and
Department of Physics, University of Crete, P.O. Box 2208, 71003
Heraklion, Greece
\label{Crete}
\and
Dept. of Physics \& Astronomy, University of California, Irvine, CA
92697, USA
\label{Irvine}
\and
School of Physics and Astronomy, Cardiff University,
  The Parade, Cardiff, CF24 3AA, UK
\label{Cardiff}
\and 
Astrophysics Group, Imperial College, Blackett Laboratory, Prince
Consort Road, London SW7 2AZ, UK
\label{Imperial}
\and
Centre for Astrophysics Research, Science and Technology Research
Centre, University of Hertfordshire, Herts AL10 9AB, UK
\label{Herts}
\and
Laboratoire d'Astrophysique de Marseille, UMR6110 CNRS, 38 rue F.
Joliot-Curie, F-13388 Marseille France
\label{Marseille}
\and
INAF - Osservatorio Astronomico di Padova,  Vicolo Osservatorio 5, I-35122
Padova, Italy
\label{Padova}
\and
Instituto de Astrof\'{\i}sica de Canarias, C/V\'{\i}a L\'{a}ctea s/n, E-38200 La
Laguna, Tenerife, Spain
\label{IAC}
\and
Departamento de Astrof{\'\i}sica, Universidad de La Laguna (ULL),
E-38205 La Laguna, Tenerife, Spain 
\label{LaLaguna}
\and
Scuola Internazionale Superiore di Studi Avanzati, via Beirut 2-4,
34151 Triest, Italy
\label{SISSA}
\and
Infrared Processing and Analysis Center, California Institute of
Technology, 770 South Wilson Av, Pasadena, CA 91125, USA
\label{IPAC}
\and
Instituto de F\'isica de Cantabria (CSIC-UC), Santander, 39005, Spain
\label{Santander}
\and
UK Astronomy Technology Center, Royal Observatory Edinburgh,
Edinburgh, EH9 3HJ, UK
\label{ROE}
\and
Institut d'Astrophysique Spatiale (IAS), B\^{a}timent 121, F-91405
Orsay, France
\label{IAS-Paris}
\and
Universit\'e Paris-Sud 11 and CNRS (UMR 8617), France
\label{Paris-Sud}
\and
Astrophysics Branch, NASA Ames Research Center, Mail Stop 245-6,
Moffett Field, CA 94035, USA
\label{NASA}
\and
Dept. of Physics and Astronomy, The Open University, Milton Keynes,
MK7 6AA, UK
\label{OpenU}
\and 
Scottish Universities Physics Alliance, Institute for Astronomy,
University of Edinburgh, Royal Observatory, Edinburgh, EH9 3HJ, UK
\label{Edinburgh2}
\and
University of Padova, Department of Astronomy, Vicolo Osservatorio
3, I-35122 Padova, Italy
\label{Padova2}
\and
Leiden Observatory, Leiden University, P.O. Box 9513, NL-2300 RA Leiden, The Netherlands 
\label{Leiden}
\and
Oxford Astrophysics, Denys Wilkinson Building, University of
Oxford, Keble Road, Oxford, OX1 3RH
\label{Oxford}
}

\date{Received 31 March 2010 / Accepted 7 April 2010}

\abstract{We use Herschel PACS and SPIRE observations of the edge-on
  spiral galaxy UGC\,4754, taken as part of the H-ATLAS SDP
  observations, to investigate the dust energy balance in this
  galaxy. We build detailed SKIRT radiative models based on SDSS and
  UKIDSS maps and use these models to predict the far-infrared
  emission. We find that our radiative transfer model underestimates
  the observed FIR emission by a factor of two to three. Similar
  discrepancies have been found for other edge-on spiral galaxies
  based on IRAS, ISO, and SCUBA data. Thanks to the good sampling of
  the SED at FIR wavelengths, we can rule out an underestimation of
  the FIR emissivity as the cause for this discrepancy. Instead we
  support highly obscured star formation that contributes little to the
  optical extinction as a more probable explanation.}


\maketitle

\section{Introduction}

Edge-on spiral galaxies are an important class of galaxies in which
the distribution and properties of interstellar dust grains can be
studied in great detail. The dust in these systems shows prominently
as dust lanes in optical images; for several edge-on spiral galaxies,
the dust distribution has been modelled by fitting realistic radiative
transfer models to such optical images \citep{1987ApJ...317..637K,
  1997A&A...325..135X, 1998A&A...331..894X, 1999A&A...344..868X,
  2004A&A...425..109A, 2007A&A...471..765B}. The conclusion of these
works is that, in general, the dust disc is thinner (vertically) but
radially more extended than the stellar disc and that the central
optical depth perpendicular to the disc is less than one in optical
wavebands, making the disc almost transparent when seen face-on.

A complementary way of studying the dust component in galaxies is through
its thermal emission at far-infrared (FIR) and submm wavelengths. A
self-consistent treatment of extinction and thermal emission, i.e.\ a
study of the dust energy balance, gives the strongest constraints on
the dust content of spiral galaxies. When we quantitatively compare
the results from extinction studies and FIR/submm emission studies, a
discrepancy is found: the relatively optically thin dust discs
emerging from the optical modelling absorb about 10\% of the stellar
radiation, whereas FIR studies of normal spiral galaxies indicate they
typically reprocess about 30\% of the UV/optical radiation
\citep{2002MNRAS.335L..41P}. This problem is clearly illustrated when
applied to individual edge-on spiral galaxies: the predicted FIR
fluxes of self-consistent radiative transfer models that successfully
explain the optical extinction generally underestimate the observed
FIR fluxes by a factor of about three \citep{2000A&A...362..138P,
  2001A&A...372..775M, 2004A&A...425..109A,
  2005A&A...437..447D}. Several scenarios (see
Sect.~{\ref{Results.sec}}) have been proposed to explain this
discrepancy, but a major problem distinguishing between these is that
the number of edge-on galaxies for which such detailed studies have
been done so far is limited, owing to the poor sensitivity, spatial
resolution, and limited wavelength coverage of the available FIR
instruments. Herschel \citep{Herschel} offers the possibility of studying
the dust energy-balance in spiral galaxies in more detail than
before. The combination of the sensitivity and the wavelength coverage
of the PACS \citep{PACS} and SPIRE \citep{SPIRE} instruments, together
covering the 70 to 500~$\mu$m region where the emission from cold dust
dominates, enables us to make reliable estimates of the total thermal
emission of the interstellar dust.
 
\begin{figure*}
\centering
\includegraphics[width=\textwidth]{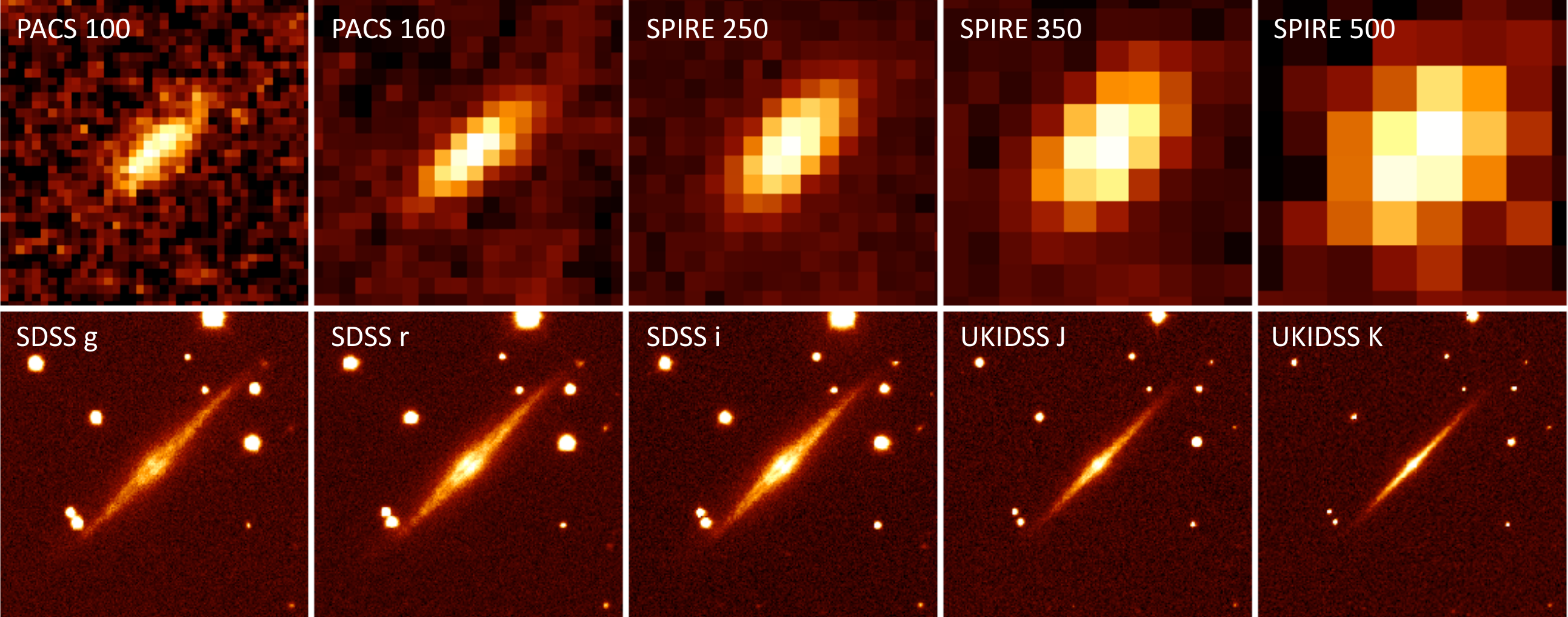}
\caption{The multi-wavelength view of UGC\,4754. The top row shows the
  Herschel images, the bottom row a selection of optical and NIR
  images from SDSS and UKIDSS. The field-of-view of each image is
  $100\arcsec \times 100\arcsec$.}
\label{Observations.pdf}
\end{figure*}


A modest number of large, nearby edge-on spiral galaxies are targetted
by Herschel key programs, including the Very Nearby Galaxies program
and the Herschel Reference Survey \citep{2010PASP..122..261B}. Much
promise, however, is offered by the Herschel Astrophysical TeraHertz
Large Area Survey \citep[H-ATLAS,][]{2009arXiv0910.4279E}, which will
image a total area of 550~deg$^2$ with PACS and SPIRE at 100, 160,
250, 350, and 500~$\mu$m. Based on the Revised Flat Galaxy Catalogue
\citep{1999BSAO...47....5K} and the 2MASS-selected Flat Galaxy
Catalogue \citep{2004BSAO...57....5M}, we expect between 100 and 200
resolved edge-on spiral galaxies in the H-ATLAS survey area.

In this Letter, we investigate the dust energy balance in the nearby
spiral galaxy UGC\,4754 as a first case. This Sbc galaxy is the
nearest \citep[$v_r=5662\pm45$~km/s;][]{2009MNRAS.399..683J} and
largest edge-on spiral galaxy in the 16~deg$^2$ field observed during
the H-ATLAS science demonstration phase (SDP). We present detailed
radiative transfer simulations of this galaxy that fit optical and
near-infrared imaging data, compute synthetic FIR fluxes and images,
and compare the model predictions to the Herschel observations at 100,
160, 250, 350, and 500~$\mu$m. In Sect.~{\ref{Observations.sec}} we
present the observations and data reduction, in
Sect.~{\ref{Analysis.sec}} we present our radiative transfer
modelling and compare the model results to the observations, and in
Sect.~{\ref{Results.sec}} we discuss these results and present out
conclusions.

\section{Observations and data reduction}
\label{Observations.sec}

The edge-on spiral galaxy UGC\,4754 was observed on 22 November 2009
with the PACS and SPIRE instruments onboard Herschel as part of the
H-ATLAS SDP observations of a $4\times4$~deg$^2$ equatorial field
centred on $(\alpha,\delta) \approx
(9^{\text{h}}05^{\text{m}},+0^\circ30\arcmin)$. Data were gathered
simultaneously in the green and red PACS bands (100 and 160~$\mu$m)
and the three SPIRE bands (250, 350, and 500~$\mu$m). The PACS and
SPIRE time-line data were reduced using {\tt{HIPE}}, with reduction
scripts based on the standard reduction pipelines.  Maps from the
SPIRE data were produced using a naive mapping technique after
removing the effects of temperature variation on the timelines using a
method developed by the H-ATLAS consortium \citep{SPIREmaps}.  Noise
maps were generated by using the two cross-scan measurements to
estimate the noise per detector pass, and then for each pixel the
noise is scaled by the square-root of the number of detector
passes. Maps from the PACS data were produced using photproject
\citep{PACSmaps}.

UGC\,4754 was clearly detected and spatially resolved in all the
observed bands (Rigby et al.\ 2010, in prep.; Smith et al.\ 2010a, in
prep.). The total fluxes are $894\pm182$~mJy at 100~$\mu$m,
$1287\pm209$~mJy at 160~$\mu$m, $878\pm134$~mJy at 250~$\mu$m,
$383\pm61$~mJy at 350~$\mu$m, and $122\pm20$~mJy at 500~$\mu$m. The
quoted errors include statistical errors, confusion noise, and a 15\%
absolute calibration uncertainty \citep{PACS,SPIRE}.

Apart from the Herschel images, we used optical {\em{ugriz}} images
from SDSS DR7 \citep{2009ApJS..182..543A} and near-infrared
{\em{YJHK}} images from UKIDSS DR5 \citep{2007MNRAS.379.1599L} for our
analysis. These images were obtained from the SDSS and UKIDSS archives
and reduced and flux-calibrated using standard reduction methods in
MIDAS. Total fluxes were measured from integrating over the
flux-calibrated maps after removal of the foreground stars. Finally,
we also used
IRAS fluxes at 60 and 100~$\mu$m from the Faint Source Catalog
\citep{1990IRASF.C......0M}.

\section{Analysis}
\label{Analysis.sec}

The first step in our analysis was to construct a radiative
transfer model for UGC\,4574, based on the {\em{gri}} and {\em{YJHK}}
images; the SDSS {\em{u}} and {\em{z}} bands were not used in our
fitting procedure because of low signal-to-noise. We used the 3D Monte
Carlo radiative transfer code SKIRT for the modelling. This code was
initially developed to investigate the effects of dust extinction on
the photometry and kinematics of galaxies \citep{2002MNRAS.335..441B,
  2003MNRAS.343.1081B}, but has evolved to a flexible radiative
transfer tool that can model the absorption, scattering, and thermal
emission of circumstellar discs and dusty galaxies
\citep[e.g.][]{2005AIPC..761...27B, 2010MNRAS.403.2053G}.

The stellar distribution was represented as a combination of a double
exponential disc (i.e.\ a disc which follows an exponential profile
both radially and vertically) and a flattened S\'ersic model. The
intrinsic stellar SED of the galaxy\footnote{To limit the numbers of
  parameters, we assumed a single intrinsic SED for the entire galaxy,
  i.e.\ we assumed no intrinsic colour gradients and the same SED for
  the bulge and disc. This assumption has virtually no effect on the
  predicted FIR emission, which is the main goal of this Letter.} was
determined by fitting a stellar population synthesis model to the
dereddened 
SDSS-UKIDSS flux densities and adapting the code used in
\citet{2008MNRAS.386.1252H, 2009MNRAS.399.1206H}; the resulting model
SED corresponds to a population of 8~Gyr old with an exponentially
decaying star formation rate and an initial burst duration of
0.15~Gyr. The dust component was also represented as a double
exponential disc. The optical properties of the dust are based on the
BARE\_GR\_S model of \citet{2004ApJS..152..211Z}, which accurately
reproduces the extinction, emission, and abundances in the Milky
Way. Our final model contains 11 free parameters: the stellar
scalelength, stellar scaleheight, bulge effective radius, S\'ersic
index, bulge flattening, bulge-to-disc ratio, total luminosity, dust
scalelength, dust scaleheight, total dust mass (or equivalently,
optical depth), and inclination.

\begin{figure}
\centering
\includegraphics[width=0.485\textwidth]{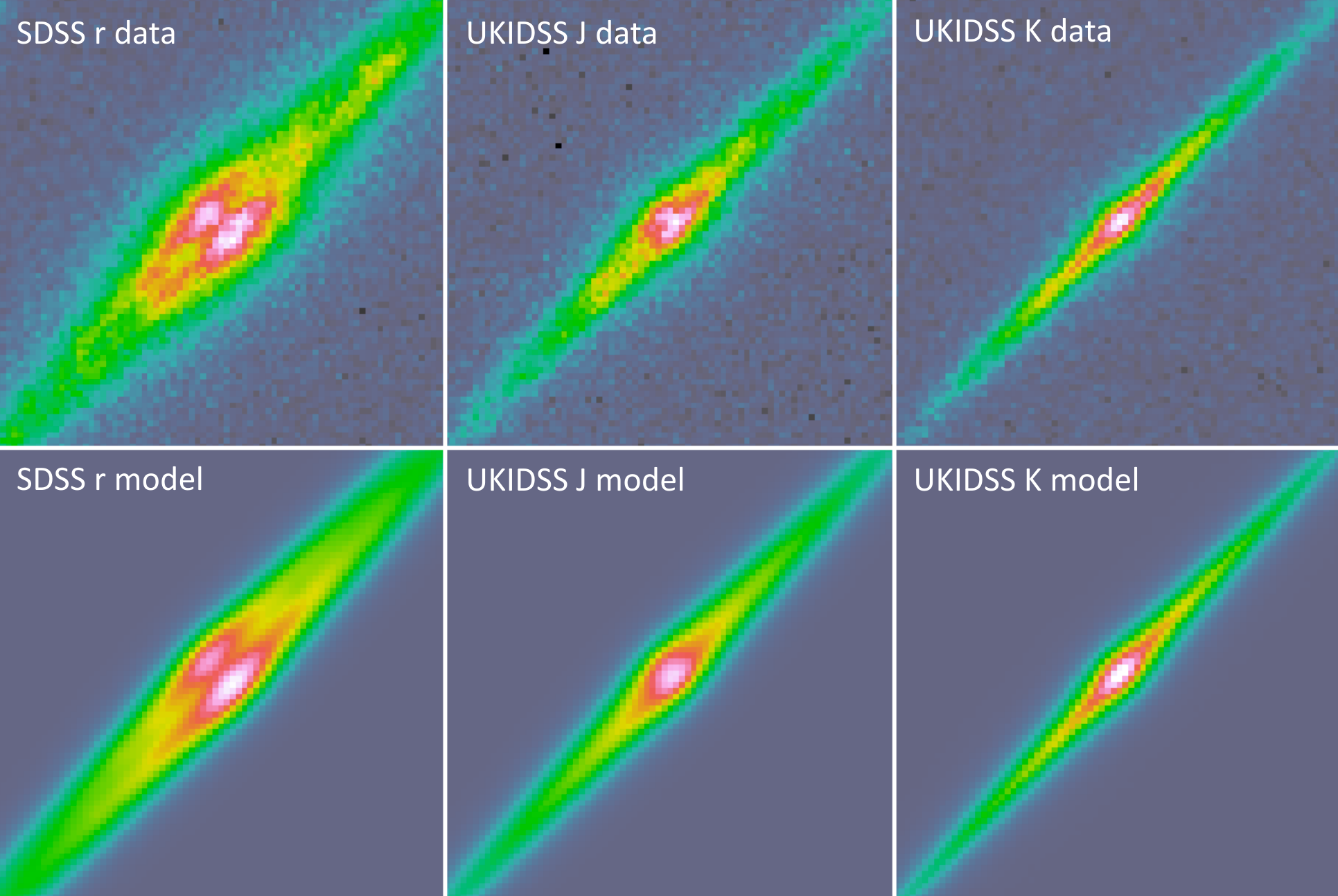}
\caption{Comparison of the observed {\em{r}}, {\em{J}} and {\em{K}}
  band images of UGC\,4754 (top row) and the results of our SKIRT
  radiative transfer modelling (bottom row). All images show only the
  central $30\arcsec\times30\arcsec$.}
\label{ModelAndData-zoom.pdf}
\end{figure}

For every choice of these 11 parameters, a set of synthetic images can
be constructed by solving the radiative transfer equation (taking both
absorption and scattering into account), convolving the resulting
radiation field with the appropriate transmission curves and
PSF. Using a $\chi^2$ minimization, we looked for the set of
parameters for which the corresponding synthetic images reproduced the
observed images best. The initial input values for the parameters of
the stellar distribution were determined from a bulge-disc
decomposition of the {\em{i}}-band image using the BUDDA code
\citep{2004ApJS..153..411D, 2008MNRAS.384..420G}. For the dust
distribution, the initial input values were calculated using the
average dust-star geometry from \citet{1999A&A...344..868X}. In
general, our fitting approach is very similar to the method applied by
\citet{1999A&A...344..868X} and \citet{2007A&A...471..765B}. The main
difference is that we fit a single photometric model to all optical
and NIR images simultaneously, whereas \citet{1999A&A...344..868X} and
\citet{2007A&A...471..765B} fit a new model to each different band.

Figure~{\ref{ModelAndData-zoom.pdf}} shows the final radiative
transfer model compared to the observed images in the {\em{r}},
{\em{J}}, and {\em{K}} bands. The stellar disc in this model has a
scalelength of 4.05~kpc and a scaleheight of 330~pc, whereas the dust
has a scalelength of 6.1~kpc and a scaleheight of 270~pc. The total
bolometric luminosity of the galaxy is $1.8\times10^{10}~L_\odot$, of
which the bulge only contributes 8\%. The total dust mass in our best-
fitting model is $1.0\times10^7~M_\odot$ and this translates into face-on
optical depths (measured along the entire $z$-axis) of 0.73, 0.49,
0.17 and 0.07 in the {\em{g}}, {\em{r}}, {\em{J}}, and {\em{K}} bands,
respectively. The corresponding edge-on optical depths (measured along
the entire line-of-sight through the centre) are 16.5, 11.1, 3.8, and
1.6. In our model, $2.1\times10^9~L_\odot$ or 12\% of the bolometric
stellar radiation is absorbed by the dust. Both the relative star-dust
geometry, with a narrow and extended dust distribution, and the modest
values of the optical depth agree with similar radiative
transfer modelling results of other edge-on spiral galaxies
\citep{1999A&A...344..868X, 2004A&A...425..109A, 2007A&A...471..765B}.

The next step in our analysis is to compare the resulting FIR
emission of our radiative transfer model to the Herschel
observations. Calculation of the FIR emission of a radiative
transfer model is carried out using SKIRT. At each position in the
galaxy, the mean intensity of the radiation field is calculated at
every wavelength during the radiative transfer simulation. From this
mean intensity, the equilibrium dust temperature of each species of
dust grains can be calculated using the energy balance equation. We
considered 10 different size bins for each of the three dust
materials in the \citet{2004ApJS..152..211Z} model (graphite, silicate,
and PAHs). We did not consider transient heating by very small
grains and PAHs in our simulation, but to estimate the impact of
transiently heated grains on the SED in the Herschel region, we
calculated the FIR emission in two ways. First, we assumed that all
grains are in LTE and we added all of their thermal emission to
calculate the FIR emission. Consequently, we repeated this, but 
only took the emission of big grains ($a>0.01~\mu$m) into account.

\begin{figure}
\centering
\includegraphics[width=0.485\textwidth]{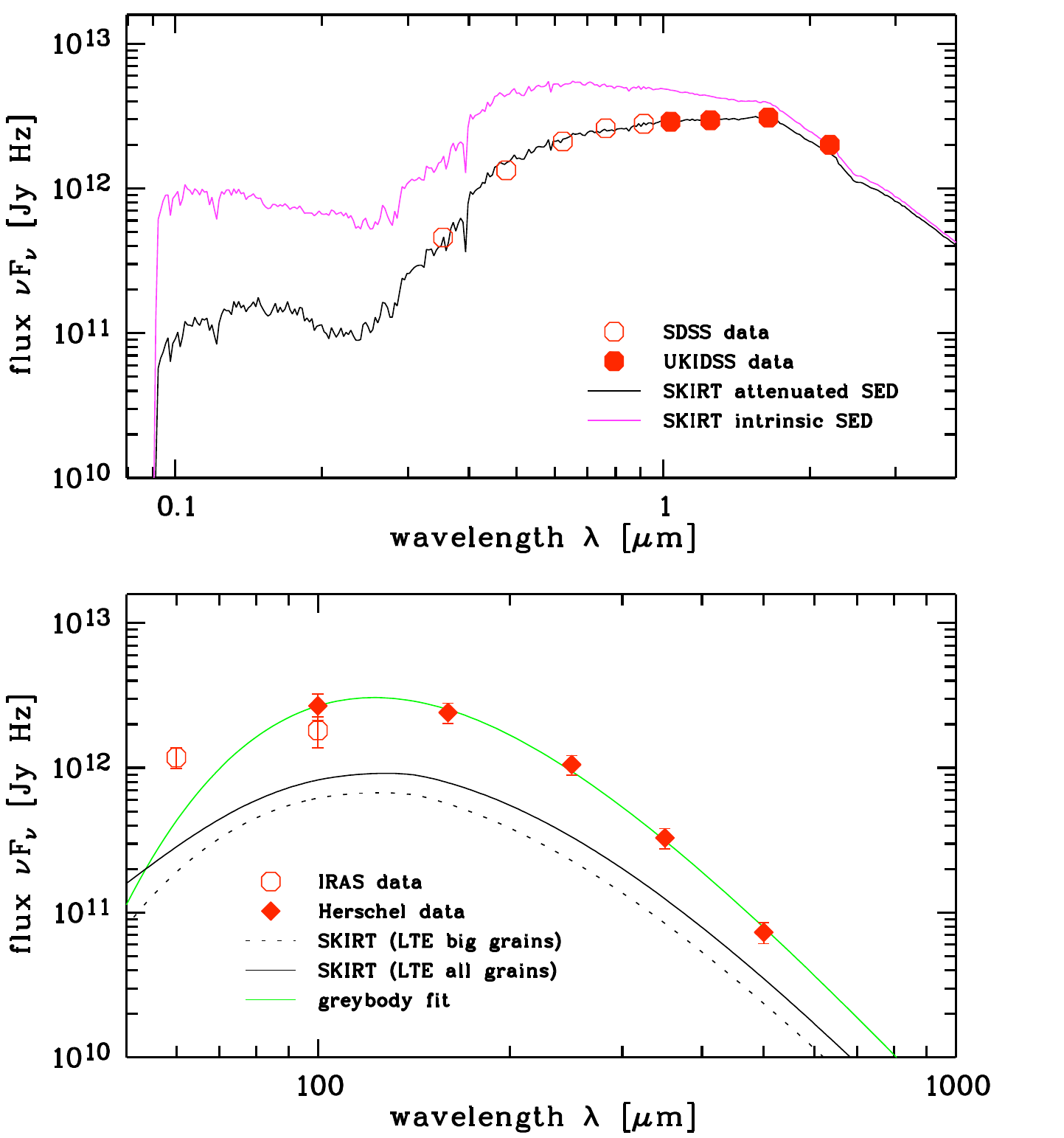}
\caption{The optical/NIR (top) and FIR/submm (bottom) spectral energy
  distribution of UGC\,4754. See label and text for the meaning of the
  different lines.}
\label{SED.pdf}
\end{figure}

Figure~{\ref{SED.pdf}} shows the spectral energy distribution (SED) of
UGC\,4754 at optical/NIR (top panel) and FIR/submm (bottom panel)
wavelengths. The solid black line in the top panel corresponds to the
attenuated SED of the SKIRT model fitted to the SDSS and UKIDSS
images, and the magenta line is the unattenuated SED. Our radiative
transfer model reproduces the optical/NIR SED very well.
In the bottom panel, the solid black line corresponds to the FIR
emission of the model assuming LTE for all grains, and the dotted line
represents the contribution of only the large grains. Our model
significantly underestimates the observed IRAS and Herschel
fluxes. This behaviour, a dust energy-balance problem, has been noted
for other spiral galaxies where the FIR emission calculated from
radiative transfer modelling is compared to the observed IRAS, ISO, 
and/or SCUBA fluxes \citep[e.g.][]{2000A&A...359...65B,
  2000A&A...362..138P, 2001A&A...372..775M}.

\section{Discussion and conclusion}
\label{Results.sec}

An apparently straightforward explanation to the dust energy-balance
problem might seem that the standard FIR emissivity, derived from dust
models finetuned to the diffuse dust emission in the Milky Way, is
actually underestimated by a factor two to three. This idea was
advocated by \citet{2000A&A...356..795A, 2004A&A...425..109A} and
\citet{2005A&A...437..447D} to explain the excess at SCUBA wavelengths
of several edge-on spiral galaxies. Support for this explanation is
the wide range of empirical values for the FIR emissivity that have
appeared in the literature \citep[see e.g.][]{1997MNRAS.289..766H,
  2004A&A...425..109A}. Thanks to Herschel's wavelength coverage
between 100 and 500~$\mu$m, we can rule out this possibility, because
we can accurately compare the absorbed stellar luminosity in our
radiative transfer model ($L_{\text{abs}} = 0.12~L_{\text{bol}}$) to
the observed luminosity $L_{\text{d}}$ emitted by the dust. We
calculated the observed dust luminosity $L_{\text{FIR}}$ in the FIR
(between 80~$\mu$m and 1~mm) by integrating the SED of the
best-fitting modified blackbody model to the Herschel data. This
model, indicated as the green line in the bottom panel of
Fig.~{\ref{SED.pdf}}, corresponds to a modified blackbody model with
$M_{\text{d}} = 1.86\times10^7~M_\odot$ and $T=19.6$~K and yields
$L_{\text{FIR}} = 5.0\times10^9~L_\odot = 0.27~L_{\text{bol}}$. Note
that $L_{\text{FIR}}$ is a lower limit to $L_{\text{d}}$, since the
mid-infrared luminosity (dominated by non-equilibrium emission by very
small grains and PAHS) has not yet been taken into account. These numbers
clearly indicate that there is a real discrepancy in the energy
balance. This discrepancy cannot be lifted by changing the FIR
emissivity at some wavelength; indeed, the energy balance equation
impies that the ratio $L_{\text{abs}}/L_{\text{d}}$ is independent of
the normalization of the emissivity.

The most likely explanation for this energy balance problem is that a
sizable fraction of the FIR/submm emission arises from additional dust
that has a negligible extinction on the bulk of the starlight. One
option is the presence of a second, thinner dust disc, originally
proposed by \citet{2000A&A...362..138P}. Such a configuration has been
successful in explaining the energy balance of spiral galaxies
\citep[][Popescu et al.\ 2010, in prep.]{2001A&A...372..775M} and the
observed attenuation-inclination relation \citep{2007MNRAS.379.1022D},
but its validity has been questioned based on deep {\em{K}} band
images of NGC\,891 \citep{2005A&A...437..447D}. An alternative hidden
dust source are young stars deeply embedded in dusty molecular
clouds. The compact dust clumps can boost the FIR/submm emission of
the dust, while keeping the extinction relatively unaltered
\citep[e.g.][]{1998ApJ...509..103S, 2000A&A...359...65B,
  2005AIPC..761..155P, 2008A&A...490..461B, 2008ApJ...672..817M}.
  
A first indication that embedded star-forming clouds might be the
solution to the case of UGC\,4574 is that the discrepancy between our
radiative transfer model and the observed FIR SED is stronger at
shorter than at longer wavelengths. This implies that warmer dust
(such as in star-forming regions) is necessary to bring the model in
balance with the data. Further evidence is the forthcoming study by
Smith et al.\ (2010b, in prep.) who analyse the SED of 1260 galaxies
from the H-ATLAS SDP field using the technique of
\citet{2008MNRAS.388.1595D}. This technique consists of fitting the
observed SED with a combination of attenuated stellar emission, dust
emission from star forming-regions, and dust emission from the general
ISM. For UGC\,4754, Smith et al.\ (2010b) find that more than 50\% of
the bolometric luminosity of the galaxy is reprocessed by dust. In
particular, their modelling predicts a significantly bluer intrinsic
UV continuum. The very efficient absorption of this radiation in
star-forming regions can power the FIR luminosity, while almost
remaining unnoticed in large-scale optical extinction maps. Combining
the results from that study with ours, we conclude that embedded star
formation is the most likely way to lift the energy balance in
UGC\,4754.

The present case study of UGC\,4754 is a demonstration of what we can
learn from detailed dust energy balance studies of edge-on spiral
galaxies. In future contributions we will apply similar modelling to
larger sets of edge-on galaxies and investigate possible systematic
links with other galaxy parameters; for example, one would expect a
correlation between the strength of the dust energy balance problem
and the star formation rate if embedded star formation is the main
driver \citep[e.g.][]{2008MNRAS.388.1595D, 2010MNRAS.403.1894D}. The
Herschel observatory and, particularly, the H-ATLAS project offer the
possibility to perform such studies in a systematic way.

\begin{acknowledgement}

  This work used data from the UKIDSS DR5 and SDSS DR7. The UKIDSS
  project is defined in \citet{2007MNRAS.379.1599L} and uses the UKIRT
  Wide Field Camera \citep[WFCAM;][]{2007A&A...467..777C}. Funding for
  the SDSS and SDSS-II has been provided by the Alfred P. Sloan
  Foundation, the Participating Institutions, the National Science
  Foundation, the U.S. Department of Energy, the National Aeronautics
  and Space Administration, the Japanese Monbukagakusho, the Max
  Planck Society, and the Higher Education Funding Council for
  England.
\end{acknowledgement}

\end{document}